\documentclass[preprint,superscriptaddress,amsmath,amssymb, floatfix,aps,pre, showpacs]{revtex4}

\usepackage{graphics,graphicx}
\usepackage{bm}

\usepackage[usenames]{color}

\begin{document}

\title{Mechanical Properties of Warped Membranes}

\author{Andrej Ko\v{s}mrlj}
\email{andrej@physics.harvard.edu}
\affiliation{Department of Physics, Harvard University, Cambridge, MA 02138}

\author{David R. Nelson}
%\email{nelson@physics.harvard.edu}
\affiliation{Department of Physics, Harvard University, Cambridge, MA 02138}
\affiliation{Department of Molecular and Cellular Biology, and School of Engineering and Applied Science, Harvard University, Cambridge, Massachusetts 02138}
%\date{\today}

\begin{abstract}

We explore how a frozen background metric affects the mechanical properties of planar membranes with a shear modulus. We focus on a special class of ``warped membranes'' with a preferred random height profile characterized by random Gaussian variables $h({\bf q})$ in Fourier space with zero mean and variance $\langle |h({\bf q})|^2 \rangle \sim q^{-d_h}$ and show that in the linear response regime the mechanical properties depend dramatically on the system size $L$ for $d_h \ge 2$. Membranes with $d_h=4$ could be produced by flash polymerization of lyotropic smectic liquid crystals. Via a self consistent screening approximation we find that the renormalized bending rigidity increases as $\kappa_R \sim L^{(d_h-2)/2}$ for membranes of size $L$, while the Young and shear modulii decrease according to $Y_R,\ \mu_R  \sim L^{-(d_h-2)/2}$ resulting in a universal Poisson ratio. Numerical results show good agreement with analytically determined exponents.

\end{abstract}

\maketitle

\section{Introduction}
\label{sec:intro}
\begin{figure*}[b]
\includegraphics[scale=.65]{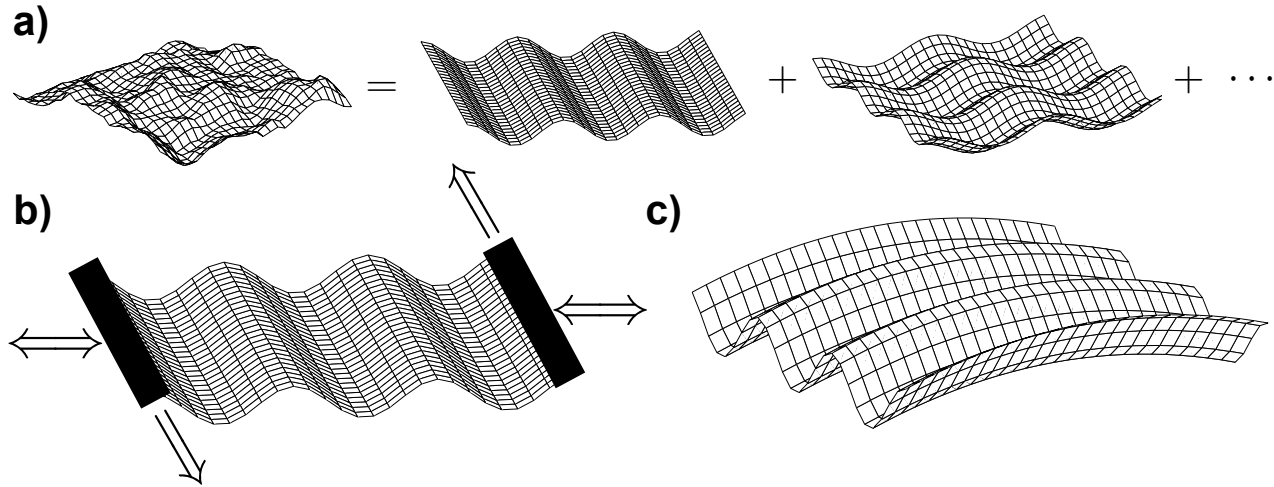}
\caption{Mechanical properties of approximately planar crumpled paper. a) The shape can be reconstructed from a linear superposition of different wave modes. b) It is easier to stretch, compress or shear in the presence of a single Fourier mode. c) However, it is harder to bend across this periodic ondulation.}
\label{fig:crumpled_paper}
\end{figure*}

Consider the mechanical properties of a thin, approximately planar piece of crumpled paper. While a flat piece of paper is almost impossible to stretch or shear, the crumpled paper can be easily stretched and sheared. On the other hand, the crumpled paper is much harder to bend, as can be observed from the response to gravity upon supporting it at only one end. Intuitively, we can understand these mechanical properties in the following way. The shape of crumpled paper can be constructed as a linear superposition of different Fourier modes (Fig.~\ref{fig:crumpled_paper}a).
It is very easy to stretch, compress or shear along the wave (Fig.~\ref{fig:crumpled_paper}b), because we are locally bending rather than stretching material, thus exploiting the stored or ``hidden'' area. At the same time bending across the wave (Fig.~\ref{fig:crumpled_paper}c) is harder due to the introduced local Gaussian curvature (nonzero radius of curvature in two directions) and the fact that it is not possible to completely release the stresses without local stretching of the material.

\begin{figure*}[b]
\includegraphics[scale=.5]{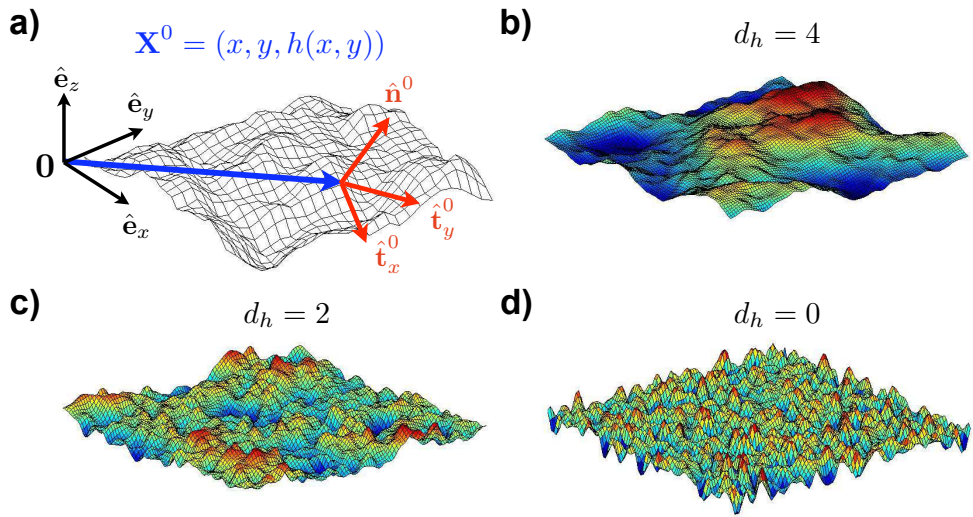}
\caption{(Color online) Warped membranes. a) The mid-plane of nearly flat warped membranes is described with a height profile $h(x,y)$ in the Monge representation. The unit vectors $\hat {\bf t}^0_x$, $\hat {\bf t}^0_y$, and $\hat {\bf n}^0$ denote local tangents and the normal to the warped membrane surface. b-d) Computer generated examples of  warped membranes characterized with different power law height distributions determined by the exponent $d_h$ (see Eq.~(\ref{eq:quenched_h_def})). For clarity, the surface heights are also indicated by a heat map, where red indicates large positive heights and blue indicates large negative ones.
}
\label{fig:warped_membrane}
\end{figure*}

In this paper we study a simplified model of the altered mechanical properties caused by a frozen background geometry. Specifically we study the effect of a quenched random set of preferred Fourier modes in a membrane with a shear modulus. Our membranes  are approximately flat and made of isotropic amorphous material with uniform thickness $t$ and whose mid-plane shape is described with a random height profile $h(x,y)$ (see Fig.~\ref{fig:warped_membrane}).
In Fourier space the height profile $h({\bf q})=\int\!d^2{\bf x} \,e^{- i {\bf q} \cdot {\bf x}} h({\bf x})/A$ is assumed to be a random Gaussian variable with zero mean and a power law variance for small $q$,
\begin{equation}
\langle |h({\bf q})|^2 \rangle = \frac{\Delta^2}{A q^{d_h}},
\label{eq:quenched_h_def}
\end{equation}
where $A$ is the membrane area, and $\langle\rangle$ denotes averaging over a quenched random Gaussian distribution funtion. The exponent $d_h$ controls the relative importance of different length scales. The case $d_h=0$ corresponds to white spatial noise where all length scales have the same amplitude, while for $d_h>0$ the amplitude of long wavelength modes is more pronounced and membranes appear ``smoother''. See Fig.~\ref{fig:warped_membrane} for surfaces generated from various probability distributions on a computer.

With the advance of 3D printing techniques it is possible to design membranes of arbitrary shapes~\cite{weaverCom}, but in principle they can be obtained experimentally as well. A whole set of individual amphiphilic bilayers could be swollen apart in a lyotropic smectic liquid crystal by adding water~\cite{larche86, safinya86}. The thermal fluctuations of the nearly flat bilayers can be described approximately by a profile $\langle |h({\bf q})|^2 \rangle \approx  k_B T / A \kappa q^4$~\cite{nelsonB},
where $k_B$ is the Boltzmann constant, $T$ the ambient temperature, $A$ the bilayer area, and $\kappa$ the bending rigidity. The fluctuating shape of the bilayer could then be ``frozen'' in by rapid polymerization~\cite{fendler84} of chemical groups embedded in the hydrocarbon chains of the amphiphilic molecules. (For an analogous polymerization experiment on spherical vesicles, see~\cite{mutz91}). The resulting profile would be described by $d_h \approx 4$. If a constant tension could be exerted on the edge of a lipid bilayer stack, or the stack oriented by application of an external electric or magnetic field prior to the polymerization, one could create surface with $d_h \approx 2$. In this case a profile due to thermal fluctuations can be described approximately by $\langle |h({\bf q})|^2 \rangle \approx  k_B T / A (\kappa q^4 + \gamma q^2)$, where $\gamma$ is related to the external tension,  or to an external electric field $E$ (magnetic field $H$) and the electric polarizability $\alpha$ (magnetic susceptibility $\chi$) of lipids with $\gamma \propto \alpha E^2$ ($\gamma \propto \chi H^2$). For sufficiently large membranes (small $q$) or strong external tensions or fields (large $\gamma$), the polymerized membranes could be described with $d_h \approx 2$. A height profile that corresponds to $d_h=0$ describes the surface of a crystal in equilibrium with its vapor below the temperature of the roughening transition~\cite{weeksB}. A frozen crystal surface of this kind could serve as a template for constructing a randomly shaped warped surface with $d_h=0$. 

In this paper we neglect thermal fluctuations and study mechanical properties for membranes characterized by $d_h =4$, $2$, and $0$ (see Fig.~\ref{fig:warped_membrane}) in the linear response regime in the presence of small forces, edge torques and external pressures across a supported membrane. Special attention is given to the asymptotic scaling of mechanical properties in the thermodynamic limit of large membrane sizes ($L$) or equivalently the long-wavelength behavior of in- and out-of-plane phonon modes (small $q$). We show that these three cases have qualitatively different behavior and that the effective in-plane elastic constants and effective bending rigidity scale with the system size $L$ for $d_h=4$, scale with the logarithm of the system size ($\ln L$) for $d_h=2$ and have no system size dependence for $d_h=0$. Only the later case agrees with expectations from conventional linear elastic theory of macroscopic materials~\cite{landauB}. 

The remainder of the paper is organized as follows. In Section~\ref{sec:free_energy} we derive the free energy cost of deformations for thin membranes of arbitrary shape by taking into account the translational and rotational symmetries. In Section~\ref{sec:shallow_shell} we derive the shallow shell equations~\cite{sanders63, koiterB} applicable for nearly flat membranes in mechanical equilibrium by minimizing the total free energy in the presence of external forces and torques. Since the shallow shell equations cannot be solved exactly, we discuss methods for solving them approximately in Section~\ref{sec:lin_response}. The iterative perturbation method (Sec.~\ref{sec:perturbation}) has a very limited practical use and fails for large membrane sizes when $d_h \ge 2$. Therefore we introduce a diagrammatic representation (Sec.~\ref{sec:diagrams}) to describe the Self Consistent Screening Approximation (SCSA) method (Sec.~\ref{sec:scsa}), which is rooted in statistical physics~\cite{ledoussal92} and enables us to determine how the anomalous elastic properties scale with system size. Finally we compare the approximate analytical results with numerical solutions of shallow shell equations (Sec.~\ref{sec:numerics}).

\section{Free energy cost of thin membrane deformations}
\label{sec:free_energy}
Deforming a nearly flat reference membrane described with a 3-dimensional position vector ${\bf X}^0 (x^k)$ parametrized by two internal parameters $x^1$, $x^2$ into configuration ${\bf X} (x^k)$ is associated with the free energy cost. In general, the free energy density $\mathcal{F}$ associated with this deformation is a function of the deformed configuration ${\bf X}$ and its derivatives $\partial_i {\bf X}$, $\partial_i \partial_j {\bf X}$,~$\ldots$~\cite{nelsonB, paczuski88, aronovitz89, guitter89, radzihovsky91}. From differential geometry~\cite{spivakB, nelsonB} we know that the first order derivatives can be interpreted as local tangent vectors ${\bf t}_i = \partial_i {\bf X}$, and the second order derivatives can be decomposed into $\partial_i \partial_j {\bf X} = b_{ij} \hat {\bf n} + \Gamma^k_{ij} {\bf t}_k$, where $b_{ij}$ is a symmetric curvature tensor, $\hat {\bf n}$ is the unit normal vector to the deformed surface, and $\Gamma^k_{ij}$ are Christoffel symbols.

The translational invariance requires that the free energy density $\mathcal{F}$ does not depend explicitly on the configuration ${\bf X}$, while the rotational invariance requires that all terms in the free energy must be scalars. For small deformations the free energy density $\mathcal{F}$ can thus be written as a ``Landau-Ginzburg''-like expansion in tangents ${\bf t}_i$ and curvatures $b_{ij}$~\cite{paczuski88, aronovitz89, guitter89}, where in general we also need to introduce couplings to the quenched fields, which could represent disorder in material or a  preferred  metric and curvatures~\cite{radzihovsky91}. For nearly flat membranes the contribution from Christoffel symbols $\Gamma^k_{ij}$ is negligible for long wavelength membrane deformations and they can be omitted in the expansion. The expansion of the free energy density in the tangents and the curvature tensor then reads
%\begin{widetext}
\begin{eqnarray}
\mathcal{F}& =& - \frac{\alpha}{2} \left({\bf t}_i \cdot {\bf t}_i\right)  + \frac{\lambda}{8} \left({\bf t}_i \cdot {\bf t}_i\right)^2 + \frac{\mu}{4} \left({\bf t}_i \cdot {\bf t}_j\right)^2 - \frac{\epsilon^S}{2} \left({\bf t}_i \cdot {\bf t}_i\right)   - \frac{\eta_{ij}^S}{2} \left({\bf t}_i \cdot {\bf t}_j\right)  \nonumber \\
&& + \frac{(\kappa-\kappa_G)}{2} b_{ii}^2 + \frac{\kappa_G}{2} b_{ij}^2 - \epsilon^B b_{ii} - \eta_{ij}^B b_{ij}
\end{eqnarray}
%\end{widetext}
where we sum over indices $i$ and $j$ and introduce position-dependent quenched fields $\epsilon^{S,B} (x^k)$ and $\eta_{ij}^{S,B}(x^k)$. We assume $\alpha>0$ which biases the system toward flat configurations and neglect quenched random disorder that is fourth order in the tangent fields and second order in curvatures. Such quenched fields could be used to describe membranes with varying thickness. 
Without loss of generality we can assume that the parameters $\eta^{S,B}_{ij}$ are traceless, i.e. $\eta^{S,B}_{ii}=0$. Terms like $b_{ij} \hat{\bf n} \cdot ({\bf t}_i \times {\bf t}_j)$ are also allowed by rotational symmetries, but they are exactly equal to $0$ due to the symmetry $b_{ij}=b_{ji}$ of the curvature tensor.

Up to an additive constant, the expansion above can be rewritten in a suggestive form,
\begin{equation}
\mathcal{F} = \frac{1}{2} \lambda u_{ii}^2 + \mu u_{ij}^2 + \frac{1}{2} \kappa K_{ii}^2 - \kappa_G \det (K_{ij}),
\label{eq:flat_f}
\end{equation}
which generalizes the well known expression for flat plates~\cite{landauB}. The first two terms represent the free energy cost of stretching and the last two terms represent the cost of bending. Above we introduced a strain tensor $u_{ij}(x^k)$ and a bending strain tensor $K_{ij}(x^k)$ given by
\begin{eqnarray}
u_{ij}& = & ({\bf t}_i \cdot {\bf t}_j - A_{ij})/2,  \nonumber \\
K_{ij}& = & b_{ij} - B_{ij} ,
\end{eqnarray}
where we introduce quenched random tensors $A_{ij} =  \delta_{ij} (\alpha + \epsilon^S)/(\mu+\lambda) + \eta^S_{ij}/\mu$ and $B_{ij}=\delta_{ij} {\epsilon}^B/(2 \kappa - \kappa_G) +  {\eta}_{ij}^B/\kappa_G$, with $\delta_{ij}$ the Kronecker delta.  For arbitrary $A_{ij}$ and $B_{ij}$, there is in general no membrane configuration ${\bf X}(x^k)$ that would correspond to the zero free energy in Eq.~(\ref{eq:flat_f}), due to geometrical frustration. A unique ground state without strains is only possible when quenched tensors satisfy the Gauss-Codazzi-Mainardi relations~\cite{spivakB} and can thus be expressed as a metric tensor $A_{ij}=\partial_i {\bf X}^0 \cdot \partial_j {\bf X}^0$ and a curvature tensor $B_{ij} = \hat {\bf n}^0 \cdot \partial_i \partial_j {\bf X}^0$ of a preferred membrane configuration ${\bf X}^0$ that corresponds to the minimum free energy. The Gauss-Codazzi-Mainardi relations can be derived from equations $\partial_2 (\partial_1 \partial_1 {\bf X^0})=\partial_1 (\partial_1 \partial_2 {\bf X}^0)$ and $\partial_1 (\partial_2 \partial_2 {\bf X^0})=\partial_2 (\partial_2 \partial_1 {\bf X}^0)$, which must be satisfied by any single valued surface.

Mechanical properties of membranes in a presence of quenched random tensors $A_{ij}(x^k)$ and $B_{ij}(x^k)$ have been studied before and it was shown that quenched averaged renormalized elastic constants can become length scale dependent~\cite{radzihovsky91, radzihovsky92, ledoussal93}. In this paper we study a particular class of quenched random tensors, which are related to each other and correspond exactly to the metric tensor ($A_{ij}$) and curvature tensor ($B_{ij}$) of the quenched random membrane configuration ${\bf X}^0(x^k)$. We call this class of quenched random surfaces, which have a unique ground state ${\bf X}^0(x^k)$ in the absence of external forces and torques, ``warped membranes''.
In the language of spin glasses these warped membranes are similar to the Mattis spin glass model~\cite{mattis76}, where the glassy ground state of spins is known, while the general case is similar to the frustrated spin glasses, e.g. the Edwards-Anderson spin glass model~\cite{edwards75}. 
 
Note that when the reference membrane configuration ${\bf X}^0(x^k)$ is \emph{not} nearly flat, the free energy density should be expressed as
\begin{eqnarray}
\mathcal{F}  = \frac{\lambda}{2} (u_i^i)^2 + \mu u_{ij} u^{ij} + \frac{\kappa}{2} (K_i^i)^2 
  + \frac{\kappa_G}{2} \left(K_{ij} K^{ij} - K_i^i K_j^j \right),
\label{eq:shell}
\end{eqnarray}
where indices are raised and lowered according to the metric tensor $g^0_{ij} = \partial_i {\bf X}^0 \cdot \partial_j {\bf X}^0$ of the reference membrane configuration ${\bf X}^0(x^k)$. This elastic description of warped membranes is known as thin shell theory~\cite{sanders63, koiterB}. If we assume that thin membranes of thickness $t$ are actually constructed from a uniform isotropic 3d material (Young modulus $E$ and Poisson ratio $\nu$), then the elastic constants in Eq.~(\ref{eq:shell}) can be expressed~\cite{sanders63, koiterB} as
\begin{equation}
\lambda = E \nu t/ (1 - \nu^2), \,\mu=E t / 2 (1 + \nu), \,\kappa = E t^3/12 (1- \nu^2), \,\kappa_G = E t^3/12 (1 + \nu).
\end{equation}
 
\section{Shallow shell equations}
\label{sec:shallow_shell}
For nearly flat warped membranes it is convenient to use the Monge representation to describe the reference warped surface
\begin{equation}
{\bf X}^0 (x,y) = x \hat {\bf e}_x + y \hat {\bf e}_y + h(x,y) \hat {\bf e}_z
\end{equation}
and then decompose the displacements of deformed membrane configuration in response to external forces into in-plane displacements $u_i(x,y)$ and out-of-plane displacements $f(x,y)$, such that 
\begin{equation}
{\bf X} = {\bf X}^0 +  u_i \hat{\bf t}^0_i + f  \hat {\bf n}^0,
\end{equation}
where $\hat {\bf t}^0_i = (\hat {\bf e}_i + (\partial_i h) \hat {\bf e}_z)/\sqrt{1 + (\partial_i h)^2}$ is a unit tangent vector and 
$\hat {\bf n}^0 = (\hat {\bf e}_z - \sum_i (\partial_i h)\hat {\bf e}_i)/\sqrt{1 + \sum_i (\partial_i h)^2}$ a unit normal vector to the warped reference surface (see Fig.~\ref{fig:warped_membrane}a). For nearly flat membranes we assume that $(\partial _i h)^2 \ll 1$ and the metric tensor $g^0_{ij} \approx \delta_{ij}$, where $\delta_{ij}$ is Kronecker's delta. Raising and lowering indices is thus a trivial operation and for simplicity we keep all indices of vectors and gradients as subscripts in the rest of the paper. In this decomposition the strain tensor $u_{ij}(x,y)$ and the bending strain tensor $K_{ij}(x,y)$ become
\begin{eqnarray}
u_{ij} & =&  \frac{1}{2} \left(\partial_i u_j + \partial_j u_i\right) + \frac{1}{2} \partial_i f \partial_j f - f \partial_i \partial_j h \nonumber \\
K_{ij} & = & \partial_i \partial_j f,
\label{eq:strain_bending_tensor}
\end{eqnarray}
where we kept only the lowest order terms. Note that the frozen spatially-varying component $h(x,y)$ breaks the inversion symmetry of $u_{ij}$ under $f \rightarrow -f$. This result is known as a shallow shell theory or the Donnell-Mushtari-Vlasov approximation~\cite{sanders63, koiterB}.

\begin{figure}[t]
\includegraphics[scale=.5]{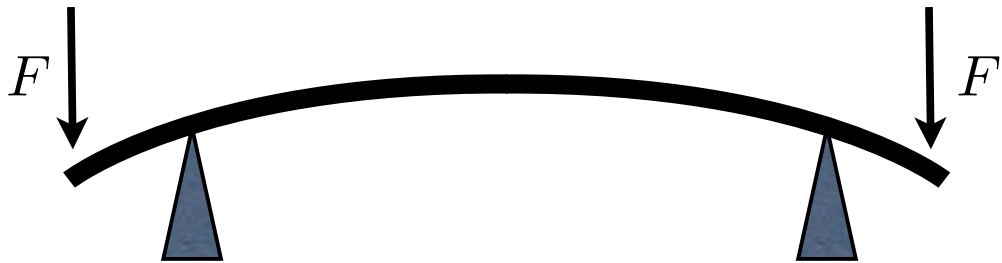}
\caption{(Color online) Edge torques acting on a membrane. The part of membrane that lies between the triangular supports, is influenced by external torques connected by a moment arm to the support. Torques are the consequence of external forces $F$ acting on the sides of the membrane that lie outside the supports.
}
\label{fig:membrane_torque}
\end{figure}

In the presence of external forces and torques the deformed membrane configuration is obtained by minimizing the total free energy functional
\begin{eqnarray}
F[u_i, f] &= &\int \! dA \  \big( \mathcal{F}  - p f \big) - \oint_{\partial A} \! ds \ (T_i u_i +  m\, \partial_n f),
\label{eq:fe}
\end{eqnarray}
where $\mathcal{F}$ is the free energy density from Eq.~(\ref{eq:shell}), $p(x,y)$ is the force/area difference across the membrane (body forces like gravity would enter in a similar way), $T_i(s)$ is an in-plane vector "tension" or force at the boundary, $m(s)$ is a local or ``edge torque'' acting at the boundary, and $\partial_n$ is the derivative in direction normal to the membrane boundary. A uniform pressure difference across the membrane would correspond to $p(x,y)\equiv p_0$. Local torques on the membrane edge can be realized by supporting the membrane and applying external force on the membrane part that lies outside the support (see Fig.~\ref{fig:membrane_torque}). In this case the membrane ``edge'' is considered to coincide with the supports. In principle terms with tangential forces inside the boundary of the membrane $- \int \! dA \, p_i u_i$ (e.g. viscous drag, electromagnetic forces) and out of plane forces at the boundary $-\oint_{\partial A} \! ds \, Q f$ could also be included, but for simplicity we assume that they are absent.

The equilibrium equations for the deformed membrane are obtained by minimizing the free energy functional in Eq.~(\ref{eq:fe}), which leads to
\begin{eqnarray}
0 & = &  \partial_j \sigma_{ij}, \nonumber \\
p & = & \partial_i \partial_j M_{ij} - \sigma_{ij} \partial_i\partial_j h - \partial_j(\sigma_{ij} \partial_i f),
\label{eq:bulk_eq}
\end{eqnarray}
where we define the stress tensor $\sigma_{ij}$ and the bending stress tensor $M_{ij}$ via 
\begin{eqnarray}
\sigma_{ij} & \equiv & \delta {\mathcal{F}}/\delta u_{ij} = \lambda u_{kk} \delta_{ij} + 2 \mu u_{ij}, \nonumber \\
M_{ij} & \equiv & \delta {\mathcal{F}}/\delta K_{ij} = (\kappa-\kappa_G) K_{kk} \delta_{ij} + \kappa_G K_{ij},
\label{eq:stress_strain}
\end{eqnarray}
At the boundary we must either prescribe displacements and slopes, $u_i$, $f$, and $\partial_n f$, or match the corresponding forces and torques
\begin{eqnarray}
T_i & = & \sigma_{ij} \hat n^{(s)}_j, \nonumber  \\
0 & = & \hat n^{(s)}_i \sigma_{ij} \partial_j f -  \hat n^{(s)}_i \partial_j M_{ij} - \hat t^{(s)}_k \partial_k \left( \hat n^{(s)}_i M_{ij} \hat t^{(s)}_j \right), \nonumber \\
m & = & \hat n^{(s)}_i M_{ij} \hat n^{(s)}_j,
\label{eq:boundary_conditions}
\end{eqnarray}
where $\hat n^{(s)}_i$ and $\hat t^{(s)}_i$ are respectively a unit normal vector and a unit tangent vector to the boundary of the membrane domain $\partial A$ in the parameter space $(x,y)$.

The equilibrium equations (\ref{eq:bulk_eq}) are complicated functions of the in-plane displacements $u_i$, and in  mechanics it is often convenient to replace them with the corresponding equations for stresses $\sigma_{ij}$. This is achieved with the introduction of the Airy stress function $\chi$, such that
$\sigma_{ij} = \epsilon_{ik} \epsilon_{j\ell} \partial_k \partial_\ell \chi$  ($\epsilon_{ij}$ is the antisymmetric Levi-Cita tensor in two dimensions) or in components
\begin{equation}
\sigma_{xx} = \partial_y \partial_y \chi,\ \sigma_{yy} = \partial_x \partial_x \chi, \sigma_{xy} = -\partial_x \partial_y \chi.
\end{equation}
Although any choice for $\chi$ yields a stress tensor that satisfies the first equilibrium equation in Eq.~(\ref{eq:bulk_eq}), the choice cannot be arbitrary. For any physically realizable stress distribution, there must correspond some displacement vector field $u_i$ such that that  the stress-strain relation in Eq.~(\ref{eq:stress_strain}) is satisfied. In plane displacements $u_i$ can be eliminated from this relation by evaluating $\epsilon_{ik} \epsilon_{j\ell} \partial_k \partial_\ell u_{ij}$, which leads to the equilibrium shallow shell equations~\cite{sanders63, novozhilovB}
\begin{eqnarray}
0 & = & \Delta^2 \chi + Y \left[ \{h, f\} + \frac{1}{2} \{f,f\} \right], \nonumber \\
p & = & \kappa \Delta^2 f - \{\chi, f\} - \{\chi, h\},
\label{eq:shallow_shell}
\end{eqnarray}
with the same boundary conditions. Above we used the Laplace operator $\Delta$ and introduced the Young's modulus $Y=4 \mu (\mu + \lambda)/(2 \mu + \lambda)$ and the Airy bracket $\{A,B\} \equiv \epsilon_{ik} \epsilon_{j\ell} (\partial_i \partial_j A) (\partial_k \partial_\ell B)$. Note that the Eqs. (\ref{eq:shallow_shell}) reduce to the familiar F\"{o}ppl-von K\'{a}rm\'{a}n equations for flat reference surfaces, i.e. for $h(x,y)\equiv0$~\cite{landauB}.

\section{Linear response mechanical properties}
\label{sec:lin_response}

In this paper we focus on the mechanical properties of warped membranes in the linear response regime in presence of small external pressure $p$, external stress $\sigma_{ij}^0$, and external torques described by the bending stress $M_{ij}^0$. In this limit the Airy stress function $\chi$ and out of plane displacements $f$ entering Eqs. (\ref{eq:shallow_shell}) are assumed to depend linearly on external forces ($p$ and $\sigma_{ij}^0$) and torques ($M_{ij}^0$) . The linear response mechanical properties of warped membrane are thus obtained by solving the linearized equilibrium equations
\begin{eqnarray}
0 & = & \Delta^2 \chi + Y \{h, f\}, \nonumber \\
p & = & \kappa \Delta^2 f - \{\chi, h\},
\label{eq:lin_shallow_shell}
\end{eqnarray}
with appropriate boundary conditions. The equations above can be solved exactly only for certain special membrane shapes (e.g. spherical caps, cylinders, etc.); therefore we treat them approximately.

As discussed above the warped membranes of interest are characterized by a unique ground state, described by a frozen height profile $h(x,y)$. In Fourier space, the $\{h({\bf q})\}$ are quenched random Gaussian variables with zero mean and  variance 
\begin{equation}
\langle h({\bf q}) h({\bf q}') \rangle = \frac{\Delta^2 \delta_{{\bf q}, - {\bf q}'}}{A q^{d_h}},
\label{eq:def_h}
\end{equation}
where $\Delta$ is an amplitude, $\delta$ is Kronecker's delta, $A$ the membrane area, and $d_h$ the characteristic exponent. Specifically we study warped membranes characterized with exponents $d_h=0$, $2$, and $4$, and determine the scaling of mechanical properties in the thermodynamic limit of large membrane sizes $L$ or equivalently in the long wavelength limit (small $q$).

First we focus on the linear response to external pressure $p\ne 0$ and no external forces ($\sigma_{ij}^0 = 0$) and torques ($M_{ij}^0 = 0$) at the membrane edge. Note from Eq.~(\ref{eq:boundary_conditions}) that $M_{ij}^0$ is related to the edge torque $m(s)$. For simplicity we assume the periodic boundary conditions. The choice of boundary condition does change the numerical prefactor of the mechanical response, but does not affect the scaling with the system size and material properties (numerical data not shown; the situation is similar to the critical force dependence on boundary conditions for the Euler buckling instability in rods~\cite{landauB}). Periodic boundary conditions enable us to rewrite equilibrium equations~(\ref{eq:lin_shallow_shell}) in Fourier space as
\begin{eqnarray}
0 & = & q^4 \chi({\bf q}) + Y \sum_{{\bf q}_1 \ne {\bf 0}} ({\bf q} \times {\bf q}_1)^2 h({\bf q} - {\bf q}_1) f({\bf q}_1), \nonumber \\
p({\bf q}) & = & \kappa q^4 f({\bf q}) - \sum_{{\bf q}_1 \ne {\bf 0}} ({\bf q} \times {\bf q}_1)^2 h({\bf q} - {\bf q}_1) \chi({\bf q}_1).
\label{eq:lin_shallow_shell_Fourier}
\end{eqnarray}
Note that we embedded the two dimensional wave vectors ${\bf q}$ in a three dimensional space by setting the third vector component to 0 in order to use the vector cross product. By solving the first equation for the Airy stress function $\chi({\bf q})$, we obtain a self-consistent integral equation for the out-of-plane displacement $f$, namely
\begin{equation}
f({\bf q}) =  \frac{p({\bf q})}{\kappa q^4}  - \frac{Y}{\kappa} \sum_{{\bf q}_1, {\bf q}_2  \ne {\bf 0}} \frac{({\bf q} \times {\bf q}_1)^2 ({\bf q}_1 \times {\bf q}_2)^2}{ q^4 q_1^4} h({\bf q} - {\bf q}_1) h({\bf q}_1 - {\bf q}_2) f({\bf q}_2).
\label{eq:integral_eq}
\end{equation}
For flat reference states ($h\equiv 0$) the external pressure of specific mode $p({\bf q})$ only induces out of plane displacements of the same mode, $f({\bf q}) = p({\bf q})/ \kappa q^4$. However, for a given realization of a nonzero warping function $h({\bf q})$, the mode couplings in Eq.~(\ref{eq:integral_eq}) also induce out of plane displacements of other modes. Nevertheless, summing the response over all possible realizations of the quenched random membrane profiles $h({\bf q})$,  leads on \emph{average} to a single induced mode, 
\begin{equation}
\langle f({\bf q}) \rangle  \equiv \frac{p({\bf q})}{\kappa_R (q) q^4},
\label{eq:kappa_R_def}
\end{equation}
where we introduced the renormalized bending rigidity $\kappa_R(q)$, and the brackets $\langle\rangle$ represent an average over reference surfaces. Note that carrying out the quenched average in Eq.~(\ref{eq:kappa_R_def}) is quite challenging, since the out-of-plane displacement $f({\bf q})$ appears on both sides of Eq.~(\ref{eq:integral_eq}).

It is important also to consider variations for different realizations of quenched random membrane profiles and ask whether they are negligible compared to the quenched average above in the thermodynamic limit. If variations are negligible, then the quenched averaged response $\langle f({\bf q}) \rangle$ for large wavelengths (small $q$) also represent the response for any given random realization of the membrane profile $h({\bf q})$ and the membranes would have the \emph{self-averaging} property. Self-averaging would imply that any given random realization of the membrane profile $h({\bf x})$ could be broken down into smaller blocks in real space, where each block would represent an \emph{independent} realization of the quenched random membrane profile. Thus the mechanical properties of the whole membrane would be characterized by the quenched averaged  properties of individual blocks. However, similar to spin glasses,~\cite{wiseman98} we expect that the quenched random membranes considered here do \emph{not} in general have the self averaging properties, because the height profiles $h({\bf x})$ have long range correlations $\langle h({\bf x}) h({\bf x}') \rangle \sim |{\bf x} - {\bf x}'|^{d_h-2}$. The only potential exception could be membranes characterized by $d_h=0$, where the height profile is completely uncorrelated $\langle h({\bf x}) h({\bf x}') \rangle \sim \delta({\bf x} - {\bf x}')$. But even in this case the numerical results discussed in Sec.~\ref{sec:numerics} imply that there is no self-averaging (see Fig. \ref{fig:bending_rigidity}, where sample to sample variations seem constant when reducing $q$ for membranes characterized by $d_h=0$).

Since the integral equation~(\ref{eq:integral_eq}) cannot be solved exactly for arbitrary reference membrane configurations $h({\bf q})$, we explain how to solve it approximately in two different regimes: a) a typical warped membrane height profile is small compared to the membrane thickness ($|h({\bf x})| \ll t$), b) the thermodynamic limit of large membrane size ($q \rightarrow 0$). The first regime can be treated with a perturbation method, where the integral equation~(\ref{eq:integral_eq}) is solved iteratively. The second regime can be approximately solved using the Self-Consistent Screening Approximation (SCSA) method, which is rooted in statistical physics~\cite{ledoussal92}. Approximations made in both regimes can be effectively understood with the diagrammatic representation. See Ref.~\cite{mezard92} for an analogous treatment of spin systems with quenched randomness.

In the next subsections we describe the iterative perturbation method, the diagrammatic representation and the SCSA method.

\subsection{Iterative perturbation method}
\label{sec:perturbation}

Our first attempt at approximately solving the integral equation~(\ref{eq:integral_eq}) is with iterative perturbation theory. The initial approximate solution for Fourier mode ${\bf q}$ of the height deviation $f({\bf x})$ is assigned as
\begin{equation}
f^{(0)}({\bf q}) =  \frac{p({\bf q})}{\kappa q^4}
\end{equation}
and the subsequent approximative solutions are constructed by iteration:
\begin{widetext}
\begin{equation}
f^{(i+1)}({\bf q}) =  \frac{p({\bf q})}{\kappa q^4}  - \frac{Y}{\kappa q^4} \sum_{{\bf q}_1, {\bf q}_2  \ne {\bf 0}} \frac{({\bf q} \times {\bf q}_1)^2 ({\bf q}_1 \times {\bf q}_2)^2}{q_1^4} h({\bf q} - {\bf q}_1) h({\bf q}_1 - {\bf q}_2) f^{(i)}({\bf q}_2).
\end{equation}
\end{widetext}
The result is a series with alternating signs, where the exact solution is reached by summing all terms in the limit $i \rightarrow \infty$. Upon averaging over all realizations of random warped membranes, each round of iteration yields a new term with an additional factor $-Y/\kappa$ and an effective height variance $h^2_\textrm{eff}(q)$, closely related to the statistics of the frozen height profile $h$. Thus we are led to a perturbation series of the form
\begin{equation}
\langle f({\bf q}) \rangle = \frac{p({\bf q})}{\kappa q^4} \left[
1-\alpha_2 \frac{Y h^2_\textrm{eff} (q)}{\kappa} + \alpha_4 \left( \frac{Y h^2_\textrm{eff} (q)}{\kappa} \right)^2 - \cdots
\right] =  \frac{p({\bf q})}{\kappa q^4} F\left( \frac{Y h^2_\textrm{eff} (q)}{\kappa} \right),
\label{eq:pert_series}
\end{equation}
where we introduced a scaling function $F$, and the $\{\alpha_i\}$  are numerical factors that can be evaluated order by order in perturbation theory. This alternating series converges only when $Y h^2_\textrm{eff} (q) / \kappa \ll 1$ or equivalently (since $Y \sim t$ and $\kappa \sim t^3$) when the typical height deviation $|h|$ of the unperturbed warped membrane is much smaller than the membrane thickness, $|h| \ll t$.

In our problem the effective height profile is
\begin{equation}
h^2_\textrm{eff}(q) \equiv \left<h^2({\bf x}; q)\right>  = \sum_{q < k < \Lambda} \langle |h({\bf k})|^2 \rangle,
\label{eq:heff_def}
\end{equation}
where $h({\bf x}; q) = \sum_{q < k < \Lambda} e^{i {\bf k} \cdot {\bf x}} h({\bf k})$ and we introduced an ultraviolet cutoff $\Lambda$ to prevent potential divergences at large momenta. One can think of this cutoff being associated with a small length scale $\Lambda^{-1}$ where the continuum description starts to break down (e.g. an atomistic scale). At long wavelengths the effective height profile scales as
\begin{eqnarray}
h^2_\textrm{eff}(q) \sim \left\{
\begin{array}{c c l}
\Delta^2 / q^{d_h - 2},  & \ \ \ & d_h > 2 \cr
\Delta^2 \ln(\Lambda/q), && d_h = 2 \cr
\Delta^2 \Lambda^{2 - d_h},  &  & d_h < 2 \cr
\end{array}
\right. .
\label{eq:heff_small_q}
\end{eqnarray}
The results above imply that for $d_h \ge 2$ the effective height profile diverges at long wavelengths (small $q$) and the alternating perturbation series in Eq.~(\ref{eq:pert_series}) does not converge! 
Therefore we need a different approach to find the asymptotic behavior of the scaling function $F$ in Eq.~(\ref{eq:pert_series}). However, first we need to introduce a diagrammatic representation. Note that for $d_h<2$ there are no divergences at long wavelengths. For this case we expect only a finite renormalization of the bending rigidity, which is $q$ independent as $q \rightarrow 0$.

\subsection{Diagrammatic representation}
\label{sec:diagrams}
\begin{figure}[t!]
\includegraphics[scale=.5]{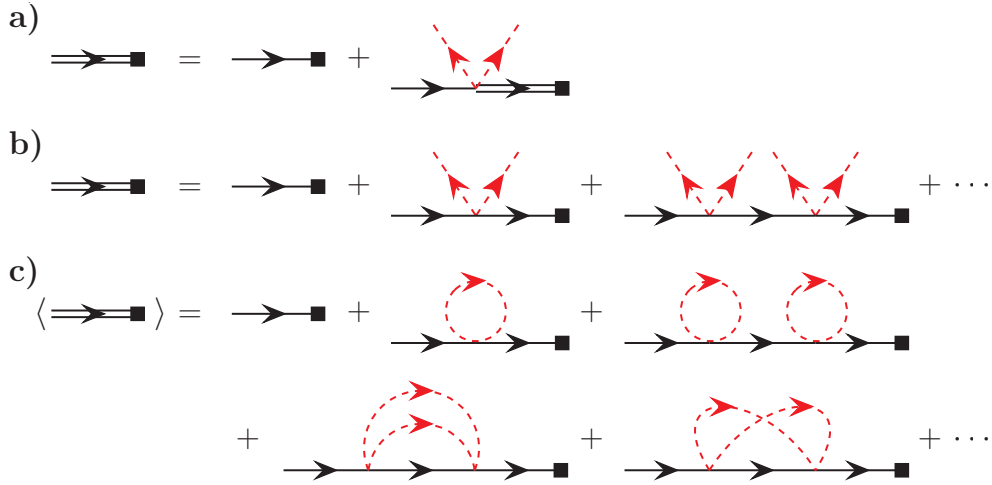}
\caption{(Color online) Diagrammatic representation of a) the integral Equation~(\ref{eq:integral_eq}), b) the iterative perturbation expansion, and c) the perturbation expansion averaged over all possible realizations of warped membranes.  Single solid lines represent propagators $1/\kappa q^4$, double solid lines represent the out of plane displacements $f$, red dashed lines represent shape profile $h$, a square represents the external pressure $p$, and each vertex carries a factor $Y$.}
\label{fig:diagrams1}
\end{figure}

The integral Equation~(\ref{eq:integral_eq}) can be schematically represented with the Feynman diagrams shown in Fig.~\ref{fig:diagrams1}.
In this scheme the ``propagator'' $1/\kappa q^4$ is represented with a black solid line, the out-of-plane displacement $f({\bf q})$ is represented with a double black solid line (indicating the ``renormalized propagator'' $1/\kappa_R(q) q^4$), each vertex carries a factor $Y$ as well as momentum factors and an even number of external legs (red dashed lines) representing $h$. The squares represent the perturbing external pressure $p$. The arrows indicate that propagators and legs carry momenta, where the sum of outgoing and ingoing momenta at each vertex must match.

Fig.~\ref{fig:diagrams1}b shows schematically the perturbation expansion, and the various approximations $f^{(i)} ({\bf q})$ to the integral equation sum only over diagrams with $i$ or less vertices. Quenched averaging over all membrane realizations is obtained by averaging every diagram over all warped membrane realizations $h({\bf x})$. Because the $\{h({\bf q})\}$ are assumed to be random Gaussian variables, Wick's theorem~\cite{amitB} allows us to average by pairing up $h$ fields (i.e. connecting red dashed legs in Fig.~\ref{fig:diagrams1}c) in all possible ways  and using the second moment averages displayed in Eq.~(\ref{eq:def_h}). In principle the exact solution is obtained by summing over all diagrams, where red dashed legs are connected in every possible way.

\begin{figure}[t!]
\includegraphics[scale=.6]{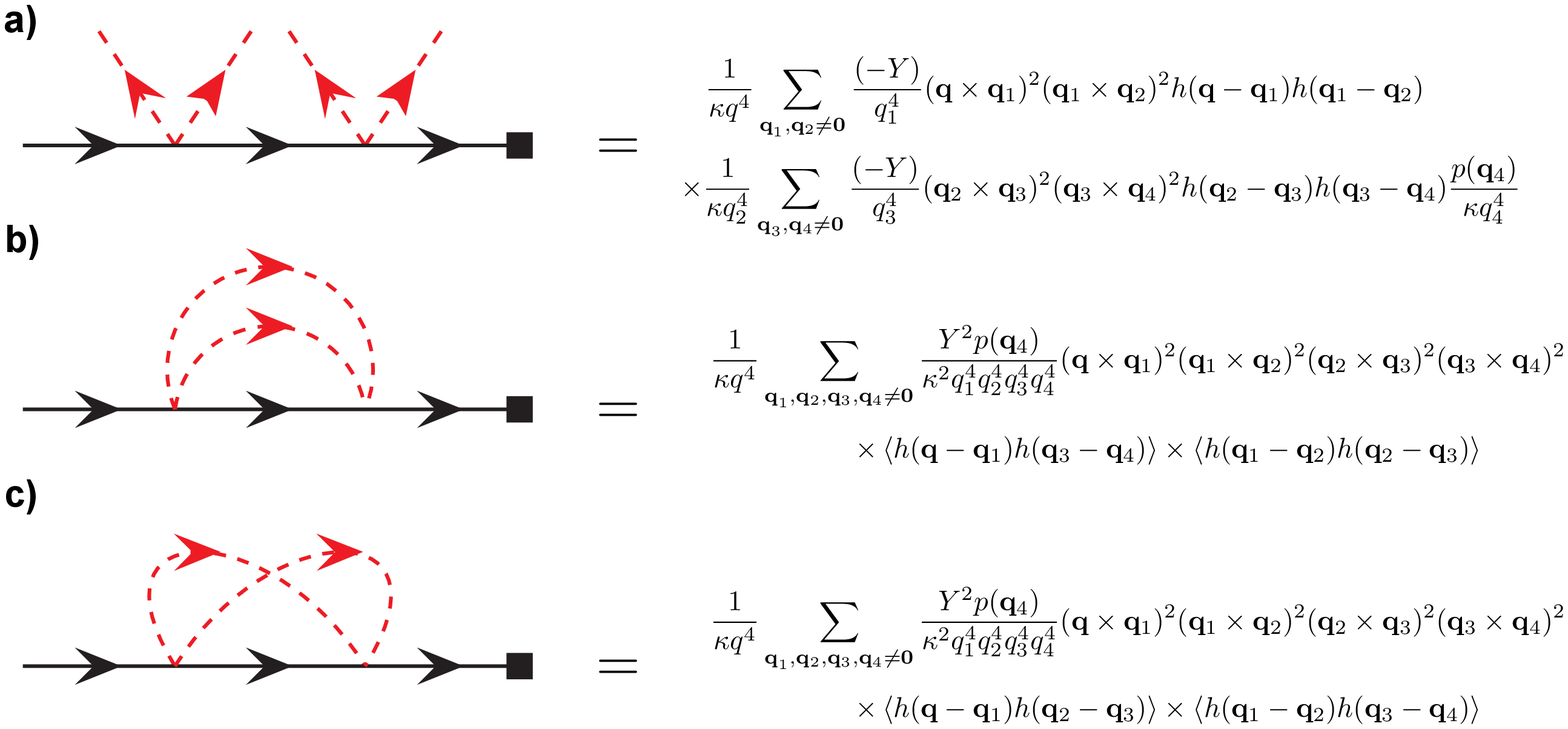}
\caption{(Color online) Algebraic expressions for sample diagrams. }
\label{fig:diagrams_key}
\end{figure}

For clarity we explicitly write the algebraic expressions for three Feynman diagrams (the first unavaraged) in  Fig.~\ref{fig:diagrams_key}.

\subsection{Self Consistent Screening Approximation method}
\label{sec:scsa}
Since it is not possible to exactly sum up all diagrams, we exploit a Self Consistent Screening Approximation  (SCSA) that sums an infinite subset of all possible diagrams to obtain the asymptotic behavior for small $q$. The SCSA method was first introduced to estimate critical exponents in the Landau-Ginzburg model of critical phenomena~\cite{bray74a, bray74b} and was later applied to calculate the effective elastic constants due to thermal fluctuations of tethered surfaces~\cite{ledoussal92, gazit09, zakharchenko10} and to study their properties in the presence of quenched random disorder~\cite{radzihovsky92, ledoussal93}. For thermally fluctuating tethered surfaces the SCSA method~\cite{ledoussal92, gazit09, zakharchenko10} gives more accurate scaling of effective elastic constants than the first order epsilon expansion in renormalization group~\cite{aronovitz88, aronovitz89}. Note that for the abstract problem, where two dimensional membranes are embedded in $d$-dimensional space the SCSA is equivalent to $1/(d-2)$ expansion and thus becomes exact when the embedding space dimension $d$ is large.

\begin{figure}[t!]
\includegraphics[scale=.45]{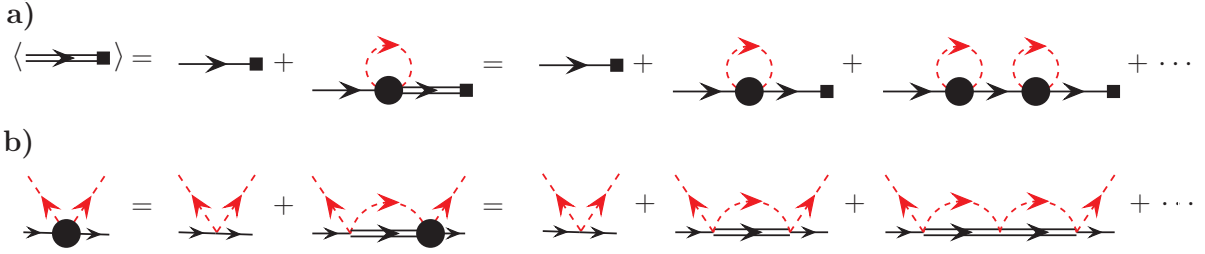}
\caption{Diagrammatic representation of the SCSA method. a) The diagrammatic series for the renormalized propagator, and b) the diagrammatic series for the renormalized vertex.}
\label{fig:diagrams2}
\end{figure}

In this paper we show how to use the SCSA to sum up all diagrams with no crossings among red dashed disordered lines (e.g. the last diagram with crossed interaction lines in Fig.~\ref{fig:diagrams1}c is excluded, while all non-crossing diagrams are included). This approximation is equivalent to expansion to the order $1/(d-2)$, while exact results at higher orders would require summing as well a subset of crossing diagrams. The SCSA infinite summation is achieved by the two self-consistent diagrammatic series in Fig.~\ref{fig:diagrams2}, where we introduce a ``renormalized vertex'' with black dot that carries a renormalized elastic coupling $Y_R({\bf q})$.
The two diagrammatic series form a geometric series, which sum up to
\begin{eqnarray}
\frac{\kappa_R({\bf q})}{\kappa} & = & 1 + \sum_{{\bf q}_1  \ne {\bf 0}} \frac{Y_R({\bf q}_1) |{\bf q} \times {\bf q}_1|^4}{\kappa  q^4 q_1^4} \langle |h({\bf q} - {\bf q}_1)|^2 \rangle \nonumber \\
\frac{Y_R({\bf q}_1)}{Y} & = &\left[1 + \sum_{{\bf q}_2  \ne {\bf 0}}\frac{Y |{\bf q}_1 \times {\bf q}_2|^4}{\kappa_R({\bf q}_2) q_1^4 q_2^4} \langle |h({\bf q}_1 - {\bf q}_2)|^2\rangle \right]^{-1}. \nonumber \\
\label{eq:scsa}
\end{eqnarray}
Note that the renormalized bending rigidity $\kappa_R({\bf q})$ is increased relative to its bare value, while the renormalized Young's modulus is always decreased. The above system of equations has to be solved self-consistently for the renormalized propagator $\kappa_R({\bf q})$ and the renormalized vertex $Y_R({\bf q})$. The coupled integral equations above and the perturbation series described earlier (Eq.~(\ref{eq:pert_series})) suggest that the renormalized quantities obey the scaling form 
\begin{eqnarray}
\frac{\kappa_R ({\bf q})}{\kappa} & = & F_\kappa \left( \frac{Y \Delta^2}{\kappa q^{d_h -2}} \right) \nonumber \\
\frac{Y_R ({\bf q})}{Y} & = & F_Y \left( \frac{Y \Delta^2}{\kappa q^{d_h -2}} \right) 
\end{eqnarray}
for membranes characterized with $d_h>2$.

It is also not possible to solve the self-consistent system of Equations~(\ref{eq:scsa}) to obtain the scaling functions $F_\kappa$ and $F_Y$ exactly. However, it is possible to obtain the self-consistent solution for the asymptotic behavior in the long wavelength limit ($q \rightarrow 0$), where we assume that
\begin{equation}
\frac{\kappa_R({\bf q})}{\kappa} = F_\kappa \left( \frac{Y \Delta^2}{\kappa q^{d_h -2}} \right) \sim C_\kappa q^{-\eta} \left(\frac{Y \Delta^2}{\kappa} \right)^{\eta/(d_h -2)},
\label{eq:scaling_kappa}
\end{equation}
where $C_\kappa$ is a constant amplitude. Dominant contributions in the sums of the self-consistent Equantions~(\ref{eq:scsa}) come from small ${\bf q}_1$ and ${\bf q}_2$, where we can use the asymptotic expressions for the renormalized quantities. This procedure leads to the remormalized vertex
\begin{equation}
\frac{Y_R({\bf q})}{Y} \sim \frac{C_\kappa q^{d_h - 2 - \eta}}{I(d_h/2, 2 - \eta/2)}  \left(\frac{Y \Delta^2}{\kappa} \right)^{(\eta - d_h + 2)/(d_h -2)},
\end{equation}
where we introduced the function $I(\alpha, \beta)$ to describe the small $q$ behavior of a sum,
\begin{eqnarray}
q^{6- 2 \alpha- 2 \beta} I(\alpha, \beta) &\equiv& \sum_{{\bf q}_1  \ne {\bf 0}}\frac{|{\bf q} \times {\bf q}_1|^4}{A q^4 q_1^{2 \alpha} |{\bf q} - {\bf q}_1|^{2 \beta}}, \nonumber \\
I(\alpha, \beta) & = & \frac{3}{16 \pi} \frac{\Gamma[\alpha + \beta - 3] \Gamma[3 - \alpha] \Gamma[3- \beta]}{\Gamma[\alpha] \Gamma[\beta] \Gamma[6- \alpha - \beta]}, \nonumber \\
\label{eq:i}
\end{eqnarray}
where $A$ is the membrane area. The renormalized vertex $Y_R$ can then be used to calculate the renormalized propagator
\begin{equation}
\frac{\kappa_R({\bf q})}{\kappa} = C_\kappa q^{-\eta} \frac{I(d_h/2, 3 + (\eta-d_h)/2)}{I(d_h/2, 2 - \eta/2)} \left(\frac{Y \Delta^2}{\kappa} \right)^{\eta/(d_h -2)}.
\label{eq:scaling_kappa2}
\end{equation}
Self-consistency requires that the asymptotic expressions in Eqs.~(\ref{eq:scaling_kappa}) and (\ref{eq:scaling_kappa2}) match, which is achieved for
\begin{equation}
\eta=\frac{d_h - 2}{2}.
\end{equation}
Thus the SCSA method predicts that in the $q \rightarrow 0$ limit renormalized quantities scale as
\begin{eqnarray}
\frac{\kappa_R({\bf q})}{\kappa} & \sim & C_\kappa q^{-(d_h -2)/2} \sqrt{\frac{Y \Delta^2}{\kappa}}, \nonumber \\
\frac{Y_R({\bf q})}{Y} & \sim & C_Y q^{+(d_h -2)/2} \sqrt{\frac{\kappa}{Y \Delta^2}},
\label{eq:ren_constants_d_g_2}
\end{eqnarray}
where $C_\kappa$ and $C_Y$ are numerical constants. Membranes characterized by $d_h=2$ have logarithmic corrections that behave like $\ln^{1/2} (\Lambda / q)$. Indeed, using the same self-consistent procedure as above we find that
\begin{eqnarray}
\frac{\kappa_R({\bf q})}{\kappa} &=&  F_\kappa \left( \frac{Y \Delta^2}{\kappa} \ln{(\Lambda/q)} \right) \sim  C_\kappa  \sqrt{\frac{Y \Delta^2}{\kappa} \ln{(\Lambda/q)}}, \nonumber \\
\frac{Y_R({\bf q})}{Y} &=&  F_Y \left( \frac{Y \Delta^2}{\kappa} \ln{(\Lambda/q)} \right) \sim C_Y  \sqrt{\frac{\kappa}{Y \Delta^2 \ln{(\Lambda/q)}}}. 
\end{eqnarray}

Interestingly, if we use the effective height profile $h^2_\textrm{eff} (q)$ introduced in the perturbation series section (Eq.~(\ref{eq:heff_small_q})) the results above can be summarized as
\begin{eqnarray}
\frac{\kappa_R({\bf q})}{\kappa} & \sim & C_\kappa \sqrt{\frac{Y h^2_\textrm{eff} (q)}{\kappa}}, \nonumber \\
\frac{Y_R({\bf q})}{Y} & \sim & C_Y \sqrt{\frac{\kappa}{Y h^2_\textrm{eff} (q)}}.
\label{eq:ren_constant_scalings}
\end{eqnarray}
Note that the SCSA method cannot be used for warped surfaces characterized by $d_h < 2$, because there is no $q$ dependence of the renormalized elastic constants in the long wavelength limit. However, we expect that the scaling description above in terms of the effective height profile to be valid for this case as well.

\subsection{Numerical results}
\label{sec:numerics}
To test our analytical results, we numerically solved the linearized shallow shell equations~(\ref{eq:lin_shallow_shell}), without making the self consistent screening approximation, using the finite difference method~\cite{numRecipesB} on a $400 \times 400$ mesh. We expect that the unfrustrated nature of the Mattis model-like unfrustrated ground state of the warped surfaces studied here allows this procedure to converge rapidly. Membrane profiles were generated in Fourier space, where $h({\bf q})=h^*(-{\bf q})$ are Gaussian random variables with zero mean and variance
\begin{eqnarray}
\langle |h({\bf q})|^2 \rangle = \left\{
\begin{array}{c c l}
\frac{\Delta^2}{A q^{d_h}} , & & q < \Lambda \cr
0 , & & q > \Lambda \cr
\end{array}
\right .
,
\end{eqnarray}
and then inverse Fourier transformed back to real space. An ultraviolet cutoff $\Lambda$ must be introduced for numerical stability, because the mesh size $a$ must be smaller than the geometrical mean of the membrane thickness $t$ and the typical curvature radius $R$, i.e. $a \ll \sqrt{R t}$. Typical values used in our numerical tests were the elastic properties of rubber with bulk Young's modulus $E = 0.1\textrm{GPa}$, Poisson ratio $\nu = 0.5$, the thickness $t = 10 \mu\textrm{m}$, the membrane size $L = 1 \textrm{cm}$, the cutoff $\Lambda$ given by $\Lambda L / 2 \pi = 8-15$, while the disorder amplitude $\Delta$ was varied systematically, such that the $h_{\textrm{eff}}(q_{\min})/t$ approximately spans a range $(10^{-1}, 10^2)$, where $q_{\min}=2\pi/L$ and $h_{\textrm{eff}}$ is defined in Eq.~(\ref{eq:heff_def}). We imposed a sinusoidal external pressure $p$ variation along one of the coordinate axes with the wave vector in the range $q L / 2 \pi = 1 - 25$ and then averaged the out-of-plane displacement response $f$ over 500 different random membrane realizations, to obtain the renormalized bending rigidity from Eq.~(\ref{eq:kappa_R_def}).

\begin{figure}[t!]
\includegraphics[scale=.5]{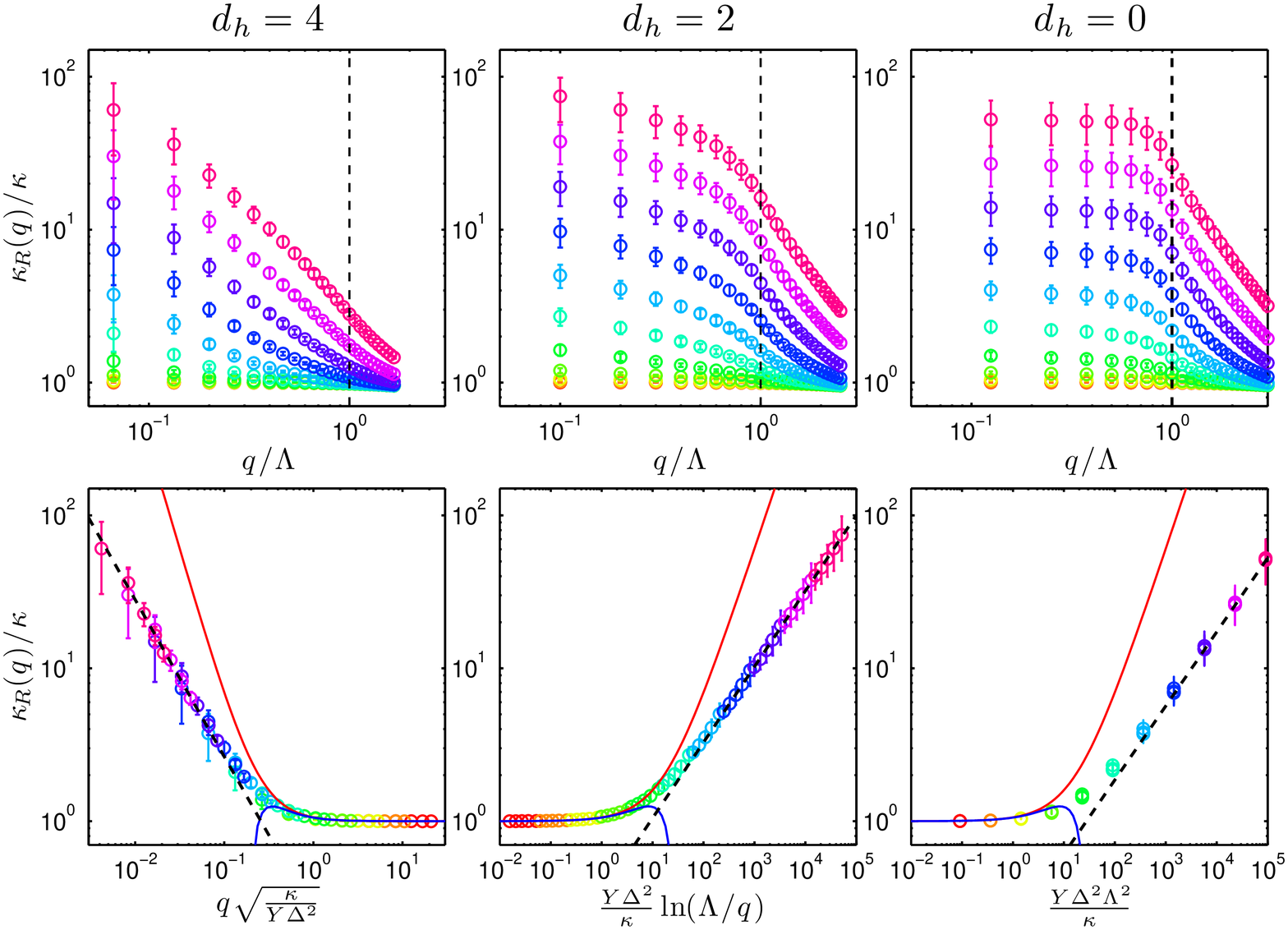}
\caption{(Color online) Numerical results for the renormalized bending rigidity $\kappa_R$ for warped membranes characterized by $d_h = 4$, $2$, and $0$. Colors of data points represent different quenched warping amplitudes $\Delta$ (violet largest, red smallest) separated by constant multiplicative factors ($\sim~3$) and errorbars  correspond to the standard deviation for different membrane samples. The black dashed lines in the first row of figures indicate the cutoff momenta $\Lambda$ of the quenched randomness. Note that $\kappa_R(q)$ is well defined even above this momentum. The second row shows high quality data collapse, where only data points with $q/\Lambda < 0.4$ were included. Black dashed lines indicate the power law fits predicted by Eqs.~(\ref{eq:ren_constants_d_g_2}-\ref{eq:ren_constant_scalings}), while the solid red and blue lines show respectively the first and second approximations obtained within the perturbation expansion.}
\label{fig:bending_rigidity}
\end{figure}

A numerical test of our theory is presented in Fig.~\ref{fig:bending_rigidity}. The first row of plots reveals two regimes: For $q \gg \Lambda$ the renormalized bending rigidity $\kappa_R$ approaches the microscopic value $\kappa$. This behavior is expected since we are trying to bend the membrane on a much smaller scale then the shortest wavelength of the quenched shape modulation, so the warped membrane locally appears flat. For $q \ll \Lambda$ we expect to observe the asymptotic scaling behavior calculated in the previous section. Indeed for membranes characterized with $d_h=4$ and $d_h=2$ the renormalized bending rigidity changes with $q$, while it levels off as $q \rightarrow 0$ for membranes characterized by white noise spatial warping with $d_h=0$. The theory also predicts that for small $q$ there is data collapse onto a single scaling function (see Eq.~(\ref{eq:pert_series})), which is shown in the second row of Fig.~\ref{fig:bending_rigidity} for data points with $q \Lambda < 0.4$. The numerically fitted scalings at small $q$ closely match the theory ($q^{-1.03}$ for $d_h=4$ and $\ln^{0.49}(\Lambda/q)$ for $d_h=2$). For $d_h=0$ we observe scaling with the combination $(Y \Delta^2 \Lambda^2/\kappa)^{0.48}$, which further supports that the asymptotic behavior for all three cases can be described with 
\begin{equation}
\frac{\kappa_R({\bf q})}{\kappa}  \sim  \sqrt{\frac{Y h^2_\textrm{eff} (q)}{\kappa}},
\end{equation}
where $h^2_\textrm{eff}(q)$ is defined in Eq.~(\ref{eq:heff_small_q}). Figure~\ref{fig:bending_rigidity} also shows that the first two terms of the perturbation series fail well before the asymptotic regime is reached.

\subsection{Linear response to external forces and torques}
Finally, we discuss the mechanical response in the presence of external forces and torques. External forces produce some average stress $\sigma^0_{ij}$ and it is common to separate out that part from the Airy stress function
\begin{equation}
\chi(x,y) = \frac{1}{2} \epsilon_{ik}\epsilon_{jl} \sigma^0_{ij} x_k x_l + \chi_r(x,y) = \frac{1}{2} \left(y^2 \sigma^0_{xx} + x^2 \sigma^0_{yy} - 2 x y \sigma^0_{xy} \right) + \chi_r(x,y).
\end{equation}
For simplicity we assume that the residual Airy stress function $\chi_r$ is still a periodic function. Shallow shell equations in the Fourier space then become
\begin{eqnarray}
0 & = & q^4 \chi_r({\bf q}) + Y \sum_{{\bf q}_1 \ne {\bf 0}} ({\bf q} \times {\bf q}_1)^2 h({\bf q} - {\bf q}_1) f({\bf q}_1) \nonumber \\
-\sigma^0_{ij} q_i q_j h({\bf q}) & = & \kappa q^4 f({\bf q}) - \sum_{{\bf q}_1 \ne {\bf 0}} ({\bf q} \times {\bf q}_1)^2 h({\bf q} - {\bf q}_1) \chi_r({\bf q}_1).
\end{eqnarray}
From the first equation above we can solve the Airy stress function $\chi_r({\bf q})$ to derive the self-consistent equation for out-of-plane displacement $f({\bf q})$, which can then be used to calculate the average in-plane strain tensor $u_{ij}^0$ created by the stress $\sigma^0_{ij}$.  Upon combining the result above with the strain tensor in Eq.~(\ref{eq:strain_bending_tensor}) and the stress-strain relation in Eq.~(\ref{eq:stress_strain}) we derive the system of equations
\begin{eqnarray}
u_{ij}^0 & \equiv & \frac{1}{A} \int \!\! dA \, \partial_i u_j = \frac{\left( \sigma^0_{ij} - \delta_{ij} \sigma^0_{kk}/2\right)}{2 \mu} +  \frac{\delta_{ij} \sigma^0_{kk}}{4(\mu + \lambda)} - \sum_{{\bf q} \ne 0} q_i q_j h(-{\bf q}) f({\bf q}), \nonumber \\
f({\bf q}) & = & - \frac{\sigma^0_{ij} q_i q_j h({\bf q})}{\kappa q_4} - \frac{Y}{\kappa} \sum_{{\bf q}_1, {\bf q}_2 \ne {\bf 0}} \frac{({\bf q} \times {\bf q}_1)^2 ({\bf q}_1 \times {\bf q}_2)^2}{q^4 q_1^4} h({\bf q}- {\bf q}_1) h({\bf q}_1 - {\bf q}_2) f({\bf q}_2).
\label{eq:average_strain_external_stress}
\end{eqnarray}

\begin{figure}[t!]
\includegraphics[scale=.5]{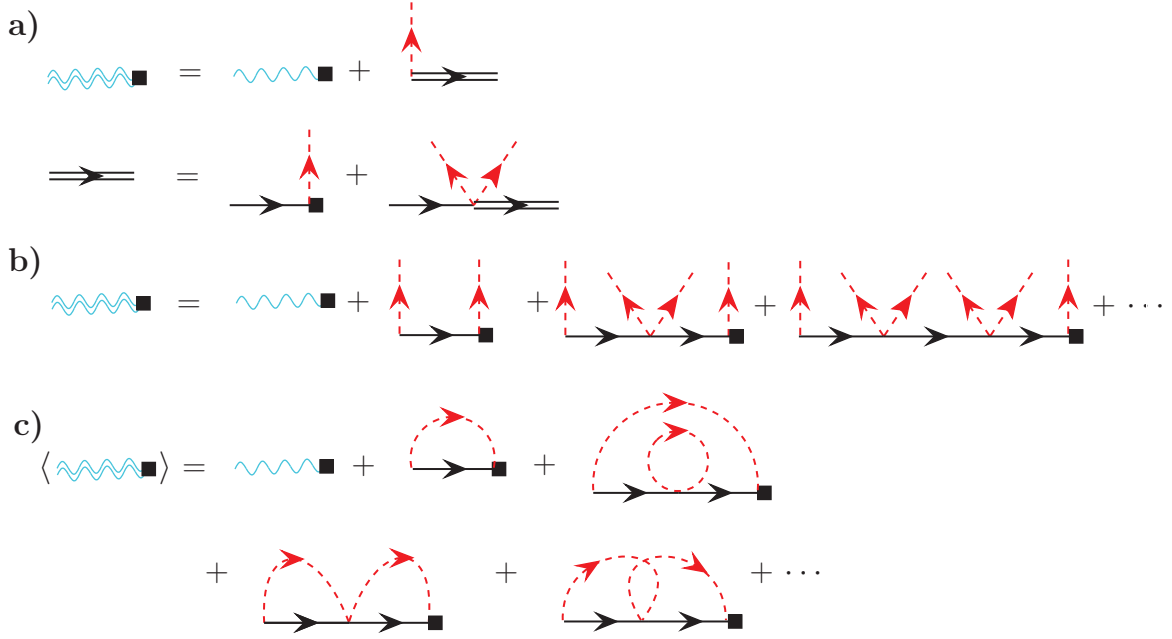}
\caption{(Color online) Diagrammatic representation of a) the system of Equations~(\ref{eq:average_strain_external_stress}), b) the associated perturbation expansion, and c) the perturbation expansion averaged over all realizations of quenched random warped membranes.  Single blue wavy lines represent the propagators $1/\mu$ and $1/(\mu+\lambda)$, double blue wavy lines represent the average in-plane-strain $u_{ij}^0$,  single solid lines represent propagators $1/\kappa q^4$, double solid lines represent the out-of-plane displacements $f$, red dashed line represent the frozen shape profile $h$, and the squares represent the external stress tensor $\sigma_{ij}^0$. Because they carry zero external wavevector, diagrams like the first one in the last row evaluate exactly to 0. }
\label{fig:diagrams4}
\end{figure}

The system of equations above can be described diagrammatically, just as in our treatment of a spatially modulated pressure (see Fig.~\ref{fig:diagrams4}).
Upon expanding the system of equations above in a perturbation series, averaging over all possible realizations of the quenched random membrane profiles $h({\bf q})$ and then summing up only the non-crossing diagrams leads to 
\begin{equation}
\langle u^0_{ij} \rangle = \frac{\left( \sigma^0_{ij} - \delta_{ij} \sigma^0_{kk}/2\right)}{2 \mu} +  \frac{\delta_{ij} \sigma^0_{kk}}{4(\mu + \lambda)} + \sum_{{\bf q} \ne {\bf 0}} \frac{q_i q_j \sigma^0_{kl} q_k q_l}{\kappa_R(q) q^4} \langle |h({\bf q})|^2 \rangle,
\end{equation}
where $\kappa_R (q)$ is the renormalized bending rigidity that we calculated above for the response to external pressure $p$.

For uniaxial stretching $\sigma^0_{xx}$ ($\sigma^0_{yy}=\sigma^0_{xy}=0$) we define the renormalized Young's modulus $Y_R$ and the renormalized Poisson's ratio $\nu_R$ by
\begin{eqnarray}
\langle u^0_{xx} \rangle & \equiv & \frac{\sigma^0_{xx}}{Y_R}, \nonumber \\
\langle u^0_{yy} \rangle & \equiv &  -\frac{\nu_R \sigma^0_{xx}}{Y_R},
\end{eqnarray}
while for a simple shear $\sigma^0_{xy}$ ($\sigma^0_{xx}=\sigma^0_{yy}=0$) we define the renormalized shear modulus $\mu_R$ by
\begin{eqnarray}
\langle u^0_{xy} \rangle \equiv \frac{\sigma^0_{xy}}{2 \mu_R}.
\end{eqnarray}

We find that the renormalized elastic constants scale as
\begin{equation}
\frac{Y_R}{Y}, \frac{\mu_R}{\mu}  \sim  \sqrt{\frac{\kappa}{Y h^2_\textrm{v}}},
\end{equation}
where we introduced the height profile variance
\begin{equation}
h^2_\textrm{v} = \langle h^2({\bf x}) \rangle = \sum_{k < \Lambda} \langle |h({\bf k})|^2 \rangle 
 \sim \left\{
\begin{array}{c c l}
\Delta^2 L^{d_h - 2},  & \ \ \ & d_h > 2 \cr
\Delta^2 \ln(L \Lambda), && d_h = 2 \cr
\Delta^2 \Lambda^{2 - d_h},  &  & d_h < 2 \cr
\end{array}
\right. .
\label{eq:profile_variance}
\end{equation}
The renormalized Young and shear modulii are thus reduced and again show power law scaling ($d_h=4$), logarithmic scaling ($d_h=2$) or no scaling ($d_h=0$) with the system size $L$. Since both renormalized elastic modulii scale in the same way with the system size for membranes characterized by $d_h \ge 2$, their ratio approaches constant value and thus a fixed universal Poisson's ratio, which is independent of microscopic material properties and is predicted to be $\nu_R=-1/3$. 

\begin{figure}[t!]
\includegraphics[scale=.5]{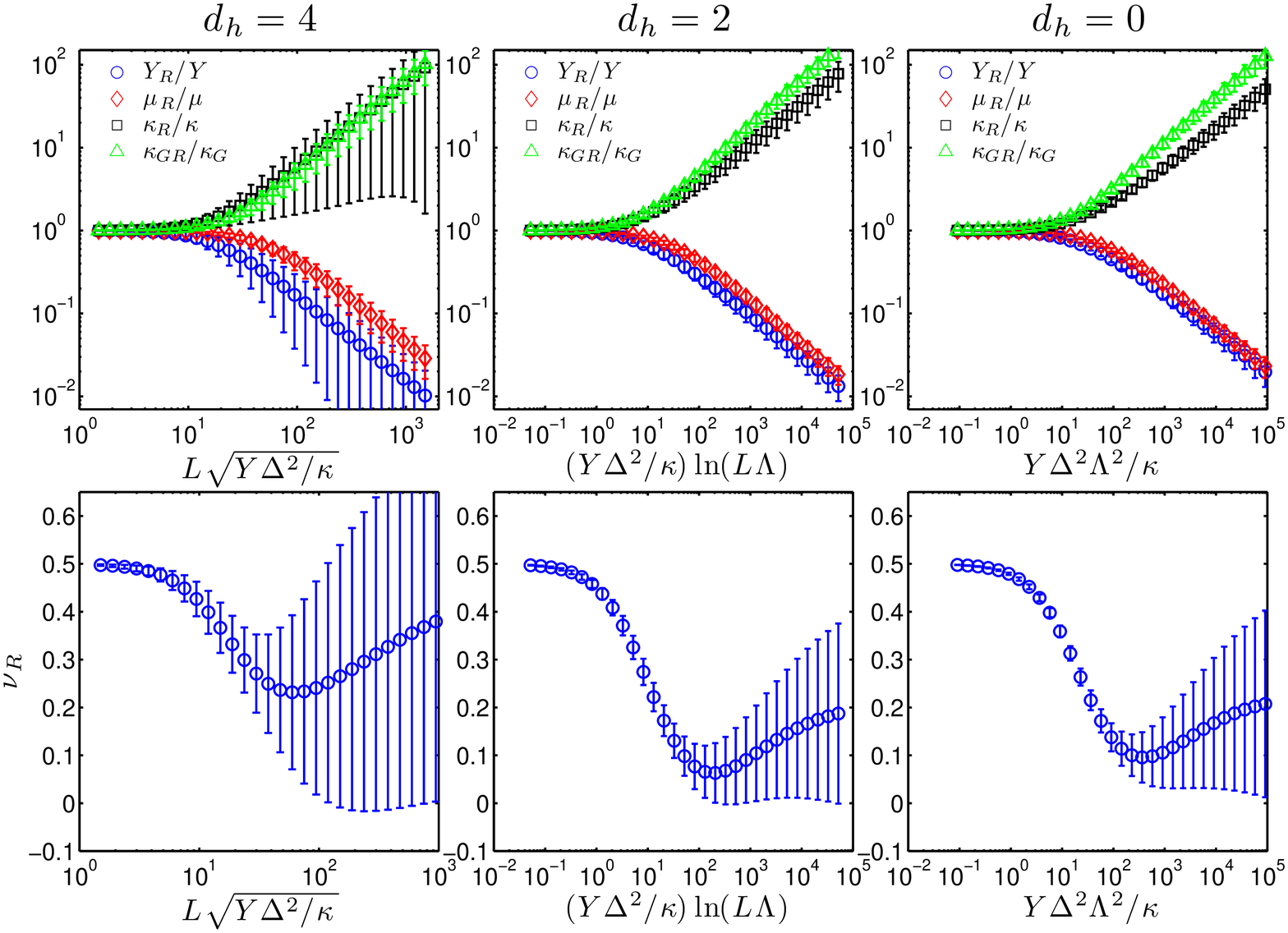}
\caption{(Color online) Numerical results for the renormalized elastic constants and Poisson ratio in the presence of external forces and torques for warped membranes characterized with $d_h = 4$, $2$, and $0$. Note that the Poisson ratio $\nu_R$ approaches its microscopic value $\nu=0.5$ for large $\kappa$ or small $L$ in all cases.  Errorbars correspond to the standard deviation for membrane samples with different realizations of the quenched random disorder.}
\label{fig:renormalized_constants}
\end{figure}

Numerical results in Fig.~\ref{fig:renormalized_constants} using the same set of parameters as above show good agreement with the predicted scaling of elastic constants ($Y_R \sim L^{-1.01}$, $\mu_R \sim L^{-1.02}$ for $d_h=4$ and $Y_R \sim \ln^{-0.49}(L)$, $\mu_R \sim \ln^{-0.52}(L)$ for $d_h=2$). However, the asymptotic value for the Poisson's ratio is quite different. This discrepancy suggests that the omitted  crossing diagrams do affect the numerical prefactors of renormalized elastic constants, while they seem to have only a small effect on the scaling exponents. Numerical results also show large sample to sample variations for $d_h=4$, where it might not be meaningful to define a fixed universal Poisson's ratio.

\begin{figure}[t!]
\includegraphics[scale=.5]{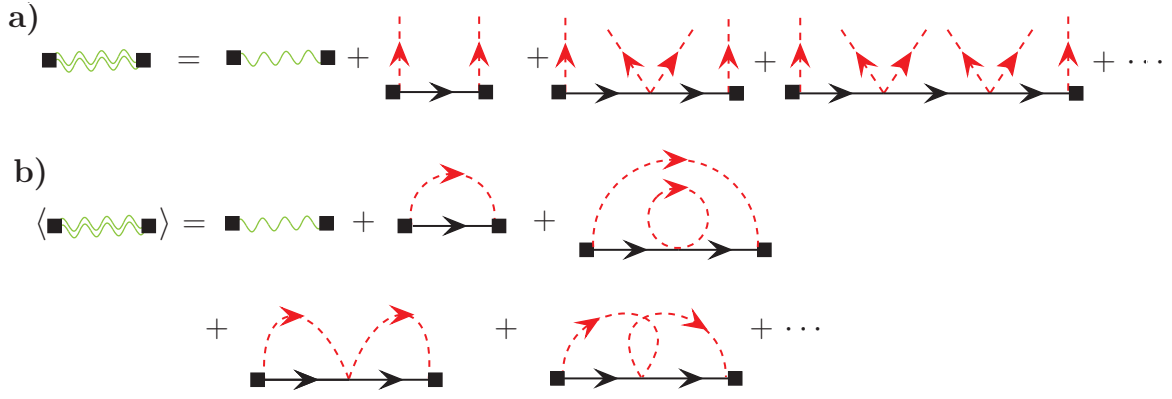}
\caption{(Color online) Diagrammatic representation of the perturbation series for the free energy cost of bending from Eqn.~(\ref{eq:free_energy_bending}), in a) unaveraged and b) averaged form. Single green wavy lines represent the propagators $\kappa$ and $\kappa_G$, double green wavy lines represent the free energy cost $F$,  single solid lines represent propagators $1/\kappa q^4$, red dashed lines represent shape profile $h$, and squares represent the average bending strain tensor $K_{ij}^0$. Note that diagrams like the first one in the last row evaluate exactly to 0. }
\label{fig:diagrams5}
\end{figure}

External torques produce some average bending strain tensor $K^0_{ij}$, which describes mean curvatures. We separate out that part from the out-of-plane displacements 
\begin{equation}
f(x,y) = \frac{1}{2} K^0_{ij} x_i x_j + f_r(x,y) = \frac{1}{2} \left(x^2 K^0_{xx} + y^2 K^0_{yy} + 2 x y K^0_{xy} \right) + f_r(x,y),
\end{equation}
and again assume that the residual function $f_r$ is periodic. The shallow shell equations in this case become
\begin{eqnarray}
Y h({\bf q}) \epsilon_{ik} \epsilon_{jl} K^0_{ij} q_k q_l & = & q^4 \chi({\bf q}) + Y \sum_{{\bf q}_1} ({\bf q} \times {\bf q}_1)^2 h({\bf q} - {\bf q}_1) f_r({\bf q}_1) \nonumber \\
0 & = & \kappa q^4 f_r({\bf q}) - \sum_{{\bf q}_1} ({\bf q} \times {\bf q}_1)^2 h({\bf q} - {\bf q}_1) \chi({\bf q}_1).
\end{eqnarray}
Calculating the average bending stress tensor $M^0_{ij}$, which is related to external torques at boundaries (see Eq.~(\ref{eq:boundary_conditions})), is the easiest by evaluating the free energy cost of macroscopically bending the membrane in Eq.~(\ref{eq:flat_f}). 
 We evaluate the free energy cost associated with bending, namely 
\begin{equation}
F/A  =  \frac{1}{2} M^0_{ij} K^0_{ij}=  \frac{1}{2} \kappa  (K^0_{ii})^2 - \kappa_G \det( K^0_{ij}) + \sum_{\bf q \ne {\bf 0}} \frac{q^4}{2} \left(\frac{1}{Y} \chi({\bf q}) \chi(-{\bf q}) + \kappa f({\bf q}) f(-{\bf q})\right).
\label{eq:free_energy_bending}
\end{equation}
Upon expanding the Airy stress function $\chi({\bf q})$ and the out-of-plane displacements $f_r ({\bf q})$ in a perturbation series and inserting them in the equation above, we arrive at a perturbation series for the free energy $F$, which can be described diagrammatically (see Fig.~\ref{fig:diagrams5})
After averaging over all possible realizations of the quenched random membrane profiles $h({\bf q})$ and then summing up only the non-crossing diagrams, we obtain
\begin{equation}
\langle F/A \rangle =  \frac{1}{2} \kappa  (K^0_{ii})^2 - \kappa_G \det( K^0_{ij}) + \sum_{\bf q \ne {\bf 0}} \frac{Y \kappa}{2 \kappa_R(q) q^4} \left(\epsilon_{ik} \epsilon_{jl} K^0_{ij} q_k q_l \right)^2 \langle |h({\bf q})|^2 \rangle,
\end{equation}
where the renormalized bending rigidity $\kappa_R(q)$ is the renormalized bending rigidity introduced before. The average bending stress tensor $\langle M_{ij}^0 \rangle$ is then
\begin{eqnarray}
\langle M_{ij}^0 \rangle  \equiv  \frac{\partial \langle F/A \rangle}{\partial K^0_{ij}} 
= (\kappa - \kappa_G) \delta_{ij} K^0_{kk} + \kappa_G K^0_{ij} + \sum_{{\bf q} \ne {\bf 0}} \frac{Y \kappa q_k q_l q_p q_r}{\kappa_R(q) q^4} K^0_{mn} \epsilon_{ik} \epsilon_{jl} \epsilon_{mp} \epsilon_{nr} \langle |h({\bf q})|^2 \rangle.
\end{eqnarray}

We define the renormalized bending rigidity $\kappa_R$ from 
\begin{equation}
\langle M^0_{xx} \rangle \equiv  \kappa_R K^0_{xx}
\end{equation}
for bending in the direction $x$ and the renormalized Gauss bending rigidity $\kappa_{GR}$
\begin{equation}
\langle M^0_{xy} \rangle \equiv  \kappa_{GR} K^0_{xy}
\end{equation}
for bending in two directions $x$ and $y$. We find that the renormalied constants scale as
\begin{equation}
\frac{\kappa_R}{\kappa}, \frac{\kappa_{GR}}{\kappa_G}  \sim  \sqrt{\frac{Y h^2_\textrm{v}}{\kappa}},
\end{equation}
where the height profile variance $h^2_\textrm{v}$ is defined in Eq.~(\ref{eq:profile_variance}). Bending rigidities are thus increased and also scale with the system size for $d_h  > 2$. Numerical results in Fig.~\ref{fig:renormalized_constants} show good agreement with the predicted scaling ($\kappa_R \sim L^{1.01}$, $\kappa_{GR} \sim L^{1.11}$ for $d_h=4$ and $\kappa_R \sim \ln^{0.50}(L)$, $\kappa_{GR} \sim \ln^{0.55}(L)$ for $d_h=2$).

Note that, as a consistency check, the free energy cost of deformations could also be used to calculate the renormalized constants $Y_R$ and $\mu_R$ in the presence of external forces and a renormalized bending rigidity $\kappa_R(q)$ in the presence of external pressure. The results are identical to the ones obtained earlier.

\section{Conclusions}
We have studied mechanical properties of nearly flat random shaped warped membranes (objects with a unique ground state in the absence of external stresses) in the linear response regime and demonstrated that when the typical height of the membrane profile is much larger than the membrane thickness $h_\textrm{v} \gg t$ the elastic constants are significantly renormalized. Using the SCSA method we found that the bending rigidities increase as $\kappa_R, \kappa_{GR} \sim \sqrt{Y h^2_\textrm{v}/\kappa}$, while the Young's modulus and shear modulus decrease as $Y_R, \mu_R \sim  \sqrt{\kappa/Y h^2_\textrm{v}}$. For membranes characterized with a random height profile $\langle |h({\bf q})| \rangle \sim q^{-d_h}$ we showed that the typical height $h_v$ scales with the system size $L$ for $d_h=4$, with the logarithm of the system size ($\ln L$) for $d_h=2$ and has no system size dependence for $d_h=0$. This leads to an anomalous system size dependence of elastic properties for membranes characterized by $d_h\ge2$.

It has been shown before~\cite{aronovitz89, guitter89, ledoussal92, gazit09, zakharchenko10} that thermal fluctuations of flat membranes also produce size-dependent renormalized elastic constants $\kappa_R, \kappa_{GR} \sim L^\eta$ and $Y_R, \mu_R \sim 1/L^{2-\eta}$, where $\eta \approx 0.85$. How thermal fluctuations affect the zero temperature scaling of elastic constants for the warped membranes discussed here will be treated in a future publication~\cite{kosmrljF}. Intuitively we expect that thermal fluctuations for membranes characterized with $d_h=0$ and $d_h=2$ lead to elastic constants that depend more strongly on system size. On the other hand, we expect that the geometric effects discussed here dominate over thermal fluctuations for membranes characterized by $d_h=4$.

Engineers are often interested in the stability of structures to compression. Since the critical stress in Euler buckling instability scales with the bending rigidity~\cite{landauB, koiterB}, we expect that the critical stress is increased for randomly warped membranes. The question remains weather the critical stress scales in the same way as the renormalized bending rigidity obtained in the linear response regime, since the non-linear terms  ignored in Eq.~(\ref{eq:shallow_shell}) drive the buckling transition. This observation suggests it would be valuable to explore the non-linear response regime as well.

\acknowledgements{We acknowledge support by the National Science Foundation, through grant DMR1005289 and through the Harvard Materials Research and Engineering Center through Grant DMR-0820484. We are also grateful to James C. Weaver and Joost J. Vlassak for discussions about how to create laboratory realizations of warped membranes.}

\bibliography{library}% Produces the bibliography via BibTeX.
\end{document}